\documentclass[aps,prapplied,reprint,superscriptaddress,showpacs,showkeys]{revtex4-2}

\usepackage{graphicx}
\usepackage{amssymb}
\usepackage{appendix} 
\usepackage{xcolor}
\usepackage[ruled,vlined]{algorithm2e}
\usepackage{float}
\usepackage{hyperref}
\usepackage{url}
\usepackage{amsmath}

\begin{document}

\title{Compact optical waveform generator with digital feedback }

\author{Shuzhe Yang}

\affiliation{University of Strasbourg and CNRS, CESQ-ISIS (UMR 7006), Strasbourg, France}

\author{Guido Masella}
\affiliation{QPerfect, 23 Rue du Loess, 67200 Strasbourg, France}

\author{Vase Moeini}
\affiliation{QPerfect, 23 Rue du Loess, 67200 Strasbourg, France}

\author{Amar Bellahsene}
\affiliation{University of Strasbourg and CNRS, CESQ-ISIS (UMR 7006), Strasbourg, France}

\author{Chang Li}
\affiliation{University of Strasbourg and CNRS, CESQ-ISIS (UMR 7006), Strasbourg, France}

\author{Tom Bienaimé}
\affiliation{University of Strasbourg and CNRS, CESQ-ISIS (UMR 7006), Strasbourg, France}

\author{Shannon Whitlock}
\email{whitlock@unistra.fr}
\affiliation{University of Strasbourg and CNRS, CESQ-ISIS (UMR 7006), Strasbourg, France}

\date{\today}

\begin{abstract}

A key requirement for quantum technologies based on atoms, ions, and molecules, is the ability to realize precise phase- and amplitude-controlled quantum operations via coherent laser pulses. However, for generating pulses on the sub-microsecond timescale, the characteristics of the optical and electronic components can introduce unwanted distortions that have a detrimental effect on the fidelity of quantum operations. In this paper, we present a compact arbitrary waveform generator that integrates a double-pass acousto-optic modulator for user-specified laser amplitude and phase modulations. Additionally, the module integrates an optical heterodyne detector to extract the precise laser pulse shape in real-time. The measured pulse shape is then fed into a digital feedback loop used to estimate the complex-valued transfer function and pre-distorted input pulses. We demonstrate the performance by generating shaped laser pulses suitable for realizing quantum logic gates with durations down to 180\,ns, requiring only a small number of feedback iterations.

\end{abstract}

\maketitle
\newpage

\maketitle
\newpage

\section{Introduction}

Quantum technologies based on the encoding and manipulation of quantum information are set to have a transformative impact in numerous areas of computation, simulation, communication, and metrology. Often this involves the coherent manipulation of quantum objects, such as atoms and ions, using shaped laser pulses. In the case of neutral atom quantum processors~\cite{ anand2024dual, graham2022multi, bluvstein2022quantum, chew2022ultrafast,evered2023high, ma2023high,bluvstein2024logical}, this is typically achieved through the electric dipole interaction between the atom and a coherent laser field where the resulting gate operation can be very sensitive to the amplitude and phase of the laser pulses~\cite{whitlock2023robust}. Quantum optimal control provides a way to design precise pulse shapes that realize desired quantum operations while minimizing the influence of noise sources~\cite{motzoi2011optimal, pagano2022error, mohan2023robust,jandura2022time, whitlock2023robust, caneva2011chopped, tovsner2009optimal, khaneja2005optimal}. However, this imposes additional constraints on the experimental hardware used to realize these pulses. Similarly, tailored laser pulses are employed in quantum simulation~\cite{scholl2021quantum, shaw2023benchmarking, omran2019generation}, optical atomic clocks~\cite{finkelstein2024universal,cao2024multi} and squeezed sensors~\cite{hines2023spin}. 

Optical pulse shaping on the sub-nanosecond timescale is widely used in ultrafast science~\cite{weiner2011ultrafast} to study the dynamics of chemical reactions and electron transfer processes with outstanding temporal resolution. Here, arbitrary optical waveforms can be realized through the use of broadband light pulses shaped in the frequency domain using dispersive optical elements~\cite{verluise2000amplitude} and spatial phase shifters~\cite{yi2011photonic}. In the radio frequency (RF) and microwave domains, arbitrary waveforms can be generated using electronic direct digital synthesizers, which are commonly used in modern radar and imaging systems, as well as in solid-state quantum computing platforms. For atomic and molecular quantum technologies, on the other hand, one typically requires visible or near-infrared optical pulses with modulation bandwidths of several megahertz. This can be achieved using acousto-optical modulators (AOMs) which diffract a continuous wave laser beam via an RF acousto-optical wave, where the amplitude, phase, and frequency of the diffracted laser beam depend on the input RF field. However, unwanted distortions~\cite{chaves2021nonlinear,rol2020time, lazuar2023calibration} in laser pulses originate from the finite propagation speed of the acoustic wave, nonlinearities of acousto-optic crystal, and limited bandwidth of the electronic devices. As the speed and accuracy of quantum operations continue to increase, even minor distortions can significantly degrade the fidelity. Therefore, it is essential to assess and mitigate hardware-induced distortions.

In this paper, we present a compact and modular arbitrary optical waveform generator capable of shaping light pulses at sub-microsecond timescales, employing closed-loop correction of the phase and amplitude distortions of the laser. Our design is based on a robust double pass AOM setup using off-the-shelf components and integrated on a small optical breadboard with a total size of $150\,\text{mm} \times 300\,\text{mm}$. The use of free-space optical components allows for high optical powers exceeding 100 mW. The system integrates an optical heterodyne detector which is used to extract the temporal amplitude and phase profile of the generated laser pulses. Moreover, for pure amplitude modulation, we observe quadrature distortion originating from the finite divergence of the laser beam interacting with the time-dependent acoustic wave. For generating a laser pulse interacting with a qubit with the desired shape, we estimate the complex-valued impulse response function using a truncated Volterra series~\cite{singh2023compensating,boyd1984analytical} and further calculate pre-distorted pulse using the Levenberg-Marquardt algorithm~\cite{more2006levenberg} to mitigate the unwanted distortions. Finally, we demonstrate the capability of our setup for generating amplitude and phase-corrected pulses with durations down to 180\,ns suitable for the realization of high-fidelity quantum gates.

\section{Optical arbitrary wave generator}

The optical setup is designed to minimize sensitivity to mechanical drifts and misalignments due to thermal expansion, while also reducing the space of the optical components. The optical mounts are based on the 1/2" Polaris system for long-term alignment stability. A schematic of the setup is shown in Fig.~\ref{fig:setup}(a). The main components are an AOM ({center frequency $100\,\text{MHz}$) in double-pass configuration and an optical heterodyne detector for beating the modulated beam against the input beam~\cite{fang2015heterodyne}. To reduce the spectral leakage~\cite{carleton1968balanced,chang2008design}, a half-wave plate is placed after the polarizing beamsplitter to rotate the polarization of the incident beam. In addition, there are beam steering mirrors for input alignment, an optional spatial filter, a three-lens telescope for mode matching the input beam to the AOM crystal, a cat eye lens for retro-reflection, and beamsplitters and a photodetector for beat detection.

\begin{figure*}
\centering
\includegraphics[width=1.0\linewidth]{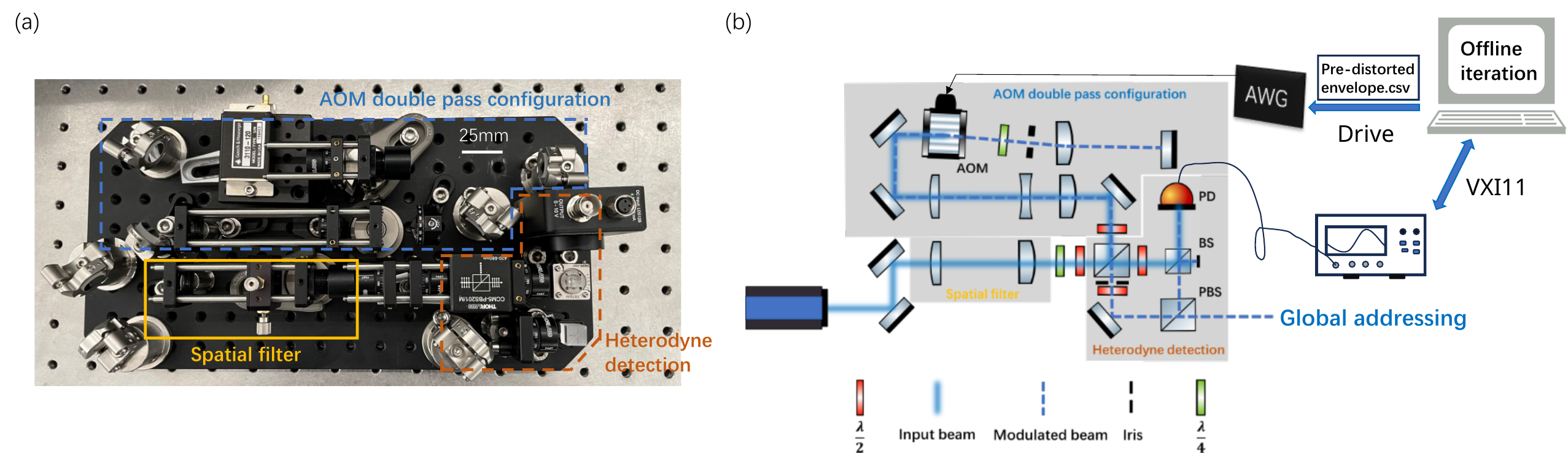}
\caption{{{Compact optical modulator and laser pulse correction feedback loop.
(a) The compact optical modulator contains three parts, AOM double pass (blue), heterodyne detection (brown), and spatial filter (yellow), (b) the beam splitter (BS) combines the two beams for generating the beat note signal, from which the temporal profile of laser's phase and amplitude are extracted. The extracted laser phase and amplitude are measured and subsequently sent to the control computer. An iterative algorithm is used to estimate the system's impulse response function. This estimated function is then used to generate a pre-distorted pulse, enabling simultaneous compensation for both amplitude and phase distortions in the laser. 
}}}
\label{fig:setup}
\end{figure*}

The three-lens configuration is designed to focus the laser beam 
into AOM to produce pulses with a short rise time while being more compact and less sensitive to input variations and misalignment than a single-lens system. In our setup, the lenses were chosen assuming an input waist of {$1200\,\mathrm{\mu m}$} and an operating wavelength of $405\,\text{nm}$. This results in a beam waist of {$63\,\mathrm{\mu m}$} on the AOM crystal while simultaneously reducing the sensitivity to input angle variations by a factor of {2.5} compared to a comparable single lens setup. We use the ABCD matrix to analyze the sensitivity to the input beam position and angle variations (see details in Appendix A). For our setup, the demagnification factor was chosen to maximize the modulation bandwidth (limited by the transit time of the acoustic wave) while keeping the Rayleigh length comparable to the crystal length. After the AOM we place a cat eye retro-reflector with a 70\,mm focal length lens and iris which allows for filtering of the $+1$ diffracted order with the widest possible modulation bandwidth.

The entire double-pass system functions as an image relay, ensuring effective mode matching of the diffracted beam with the input beam. The good alignment maximizes contrast at the optical heterodyne detector, which utilizes a Mach-Zehnder interferometer coupled with a fast photodetector (Thorlabs PDA015A2). The photodetector is electronically isolated from the mechanical components to minimize noise and improve signal quality.

To demonstrate the robustness of the compact setup, we tested the pointing stability as a function of environmental temperature in the range from 297\,K to 305\,K under continuous wave operation. For comparisons, we also tested a single lens double-pass AOM setup which is typical of an atomic, molecular, and optical physics laboratory, used for reference, see Appendix A. For angular accuracy, we find a thermal coefficient of $10\,\mu\text{rad/K}$, and for position we find $2\,\mu\text{m/K}$. In contrast, the reference (lab-based) setup exhibits $23\,\mu\text{rad/K}$ and $11\,\mu\text{m/K}$ respectively. We attribute the improvement to the three-lens configuration design for which the position and pointing are less sensitive to input angle deviations than the comparative setup, according to an ABCD matrix analysis of the three-lens design presented in Appendix A. Angular stability is particularly important, as it strongly affects the diffraction efficiency of the AOM. According to simulations based on the measurement data at different temperatures, as shown in Appendix A, we find that the compact setup can tolerate up to a $9 \,K$ change, whereas diffraction is almost completely lost for the reference setup with $4 \,K$ temperature change. 

\section{Pulse shape characterization}

A key feature of the presented optical Arbitrary Waveform Generator (AWG)~\cite{scott2010dynamic, cundiff2010optical} is its ability to characterize the generated pulses in situ using a heterodyne detector. The modulated pulses after double-pass AOM are partially mixed with the unmodulated optical reference and thus a beat note is created at twice AOM's operating frequency. This beat is detected by a fast photodetector and then recorded on a digital oscilloscope.

We use the digital IQ demodulation method to extract the amplitude and phase of the time-dependent laser pulse from the beat note signal. This involves a zero-phase Butterworth low-pass filter implemented with second-order sections and forward-backward filtering~\cite{virtanen2020scipy}, which minimizes edge distortion. We confirm via numerical simulations that extract a $2\,\mathrm{MHz}$ temporal profile from $200\,\mathrm{MHz}$ beat note signal performs better than finite impulse response filters, with relative errors $<10^{-4}$ characterized by Mean Absolute Scaled Error (MASE) (Eq.~\eqref{eq:MASE}) between the real pulse shape and the demodulated pulse shape for pulse durations down to 100\,ns.

{In Fig.~\ref{fig:distortion}, we show a simple pulse generated and characterized by our optical AWG. The input waveform is a truncated Gaussian pulse (pure amplitude modulation) with a full width at half maximum of $0.3\,\mu\mathrm{s}$. The measured amplitude of the optical pulse has a similar Gaussian shape, with a time delay of $1.4\,\mu\mathrm{s}$ expected for the propagation time of the acoustic wave in the AOM crystal, as shown in Fig.~\ref{fig:distortion}(b). The actual laser amplitude shape experienced by the atom can be distorted due to various factors, including dispersion and nonlinearities introduced by the finite bandwidth and impedance mismatching of the electrical components between the AWG and the AOM. The resulting laser amplitude shape it imprints can be mathematically expressed as~\cite{singh2023compensating,boyd1984analytical}:

\begin{align}\label{eq:volterra}
A^{\text{out}}(t) &= h_0 + \sum_{n=0}^{\infty} \int_{-\infty}^{\infty} \dots \int_{-\infty}^{\infty} h_n(t-\tau_1, \dots, t-\tau_n)\cdot \notag \\
&\quad \prod_{i=1}^{n} A^{\text{in}}(\tau_i) \, d\tau_1 \dots d\tau_n
\end{align}

\noindent
where $A{(t)}$ is the amplitude temporal profile of the laser pulse, $h_0$ is the zero-order term, and 
$h_n(t-\tau_1, \ldots, t-\tau_n)$ are the nth-order Volterra kernels (impulse response functions) of the system. This equation represents the full Volterra series expansion of the nonlinear system.

In general, the Volterra kernels can be complex-valued, leading to quadrature distortion. This effect manifests in our system as an undesired phase modulation $\phi'(t)$, exhibiting a peak amplitude of approximately 0.25 rad. Thus, if left uncorrected, this pulse would have a detrimental effect on the fidelity of quantum operations as the rotation axis of this quantum operation on the Bloch sphere, is rotated by $\phi'(t)$ around $z$ axis throughout the pulse~\cite{mckay2017efficient}.

Quadrature distortion, as seen in Fig.~\ref{fig:distortion}(a), is an intrinsic property of the acousto-optic interaction with a focused beam. To understand its origin, we consider an acoustic wave with amplitude temporal profile $P(t)$ traveling through the AOM crystal along $z$ with velocity $v$ and central acoustic wave frequency $f$. As $f$ is typically much smaller than the optical frequency, we can regard the refractive index inside the AOM as a slowly varying sinusoidal function in space, $n(z,t) = n_0 - \eta P(t-z/v)\cos(qz-\phi)$, where $q$ is the wavenumber of acoustic wave and $\eta$ (the change of the refractive index) is proportional to the acoustic wave amplitude.

\begin{figure*}
\includegraphics[width=1.0\linewidth]{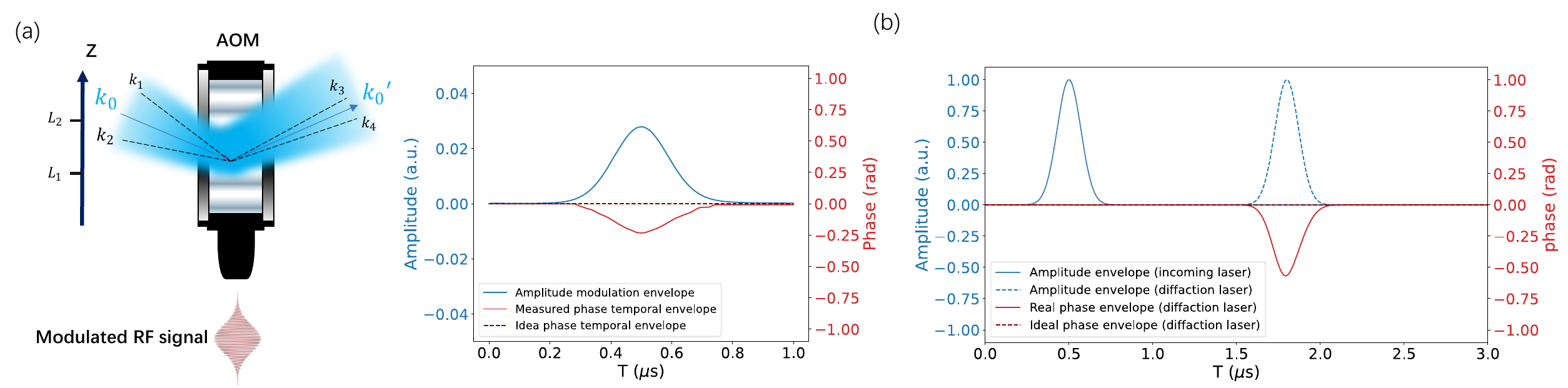}
\caption{\label{fig:distortion}{The phase distortion induced from AOM. (a) To generate fast laser pulses with short rise time by AOM, we focus a laser beam with central wavevector $k_0$ traveling through AOM. The diffraction light with wavevector $k_0'$ results from interaction between incoming laser beam and the acoustic wave which is modulated by $P(t)$ with wavenumber $q$ driving AOM 
crystal, $k_1$ and $k_2$ represent the range of wavevectors present in the focused Gaussian beam due to its angular spectrum, while $k_3$ and $k_4$ are the corresponding range of wavevectors in the diffracted beam that satisfy the tolerance of AOM Bragg condition. The phase of the laser is extracted from the cases of pure AOM amplitude modulation, and the corresponding analytical simulation based on Eq.~\eqref{eq:quadraturephase} is shown in (b).}}
\end{figure*}

To calculate the complex amplitude reflectance of the light in the first-order diffracted beam, we could divide the medium into incremental planar layers orthogonal to the z-axis. The incident optical plane wave is partially reflected at each layer because of the refractive-index change~\cite{saleh2019fundamentals}. 
We assume that the gradient of the refractive index is sufficiently small so that the transmitted light between layers is conserved as it propagates, and the Bragg angle $\theta_0$ is small. Thus, we obtain an expression of total complex amplitude reflectance being the sum of all incremental reflectance in the interaction region~\cite{saleh2019fundamentals}:

\begin{equation}\label{eq:reflectance}
 r = -i  
 r_0e^{i\phi}\int_{L1}^{L2} P(t-z/v)e^{i(2k(z)\sin\theta(z)-q)z}dz,
\end{equation}

\noindent
 where $r_0=\frac{q \Delta n}{4n(\sin^2\theta)}$ and $q=k_0\sin\theta_0$, $L_{1}$ and $L_{2}$ indicates the interaction region of AOM crystal and laser beam. For a collimated beam with wavevector $k(z)=k_0$ the maximum diffraction efficiency occurs at the Bragg angle $\theta=\sin^{-1} (q/2k_0)$. However, in practice, the small waist of the laser beam results in a spread of momenta $k(z)$. Thus plane wave components slightly away from the Bragg angle can also be reflected, combined with the finite divergence of acoustic wave traveling through the AOM, resulting in a non-zero term $2k(z)\sin\theta(z)-q\ne 0$. To determine the time-dependent phase shift induced by the AOM on the diffracted beam, we extract it from the complex amplitude reflectance with (neglecting the phase offset $\pm \frac{\pi}{2}$)(Fig.~\ref{fig:distortion}(b)):
\begin{equation}\label{eq:quadraturephase}
\theta'(t) = \textrm{arccos}[\textrm{Re}(r)]
\end{equation}

For pure amplitude modulation, the phase of the laser pulse acquires an additional time-varying phase whose temporal profile is related to the RF amplitude temporal envelope, consistent with what we observed experimentally (Fig.~\ref{fig:distortion}(a)). Consequently, for any laser pulse with a time-varying amplitude envelope generated by an AOM, the corresponding induced phase shift can result in unexpected infidelity in quantum operations.

\section{pulse correction with digital feedback}

If the system's transfer function can be accurately determined, it should be possible to compute a pre-distorted input signal that results in the desired output pulse, for example, through deconvolution. However, in practice, deconvolution is numerically unstable for noisy data and cannot fully account for non-linear distortion. Additionally, the transfer function may drift over time depending on environmental conditions. Therefore, we employ an iterative feedback loop to efficiently estimate the system's effective transfer function and compute a corresponding pre-distorted input pulse, enabling us to correct the distortions in laser pulses.

Our optical AWG modulates laser pulses using a double-pass AOM driven by RF fields. The relatively slow amplitude and phase modulation of the RF fields determine the temporal shape of the laser pulses. The complex-valued input (target) and output (measured) envelope~\cite{gustavsson2013improving,wittler2021integrated} of the laser can be expressed in terms of its amplitude $A^{\alpha}(t)$ and phase $P^{\alpha}(t)$: 

\begin{equation}\label{eq:pred}
s^{\alpha}(t) = I^{\alpha}(t) + iQ^{\alpha}(t),
\end{equation}

\noindent
where $\alpha \in \{\text{in}, \text{out}\}$, $I^{\alpha}(t)=A^{\alpha}(t)\cos[P^{\alpha}(t)]$ and $Q^{\alpha}(t)=A^{\alpha}(t)\sin[P^{\alpha}(t)]$.

We characterize the effective transfer function by using the truncated Volterra series method~\cite{singh2023compensating, mathews2000polynomial}. In a time-invariant, causal system, the output signal can be expressed as the sum of the $N$ nonlinear elements (Eq.~\eqref{eq:volterra}), the Volterra kernel in each element can be considered as the higher-order impulse response of the system. In our case, we found that the impulse response function estimated by using the Volterra series truncated to first order is sufficient to reproduce the output laser amplitude and phase shape for a given pulse, which we refer to as ``fitted output". According to the discretized form of the signal (Eq.~\eqref{eq:pred}) obtained from measurement, the relationship between input and output signal can be expressed as:

\begin{equation}\label{eq:vl}
    s_{n}^{out} = h^{(0)} + \sum_{j=0}^{R-1} h_j^{(1)} s_{n-j}^{in}, 
\end{equation}

\begin{figure*}
\includegraphics[width=1.0\linewidth]{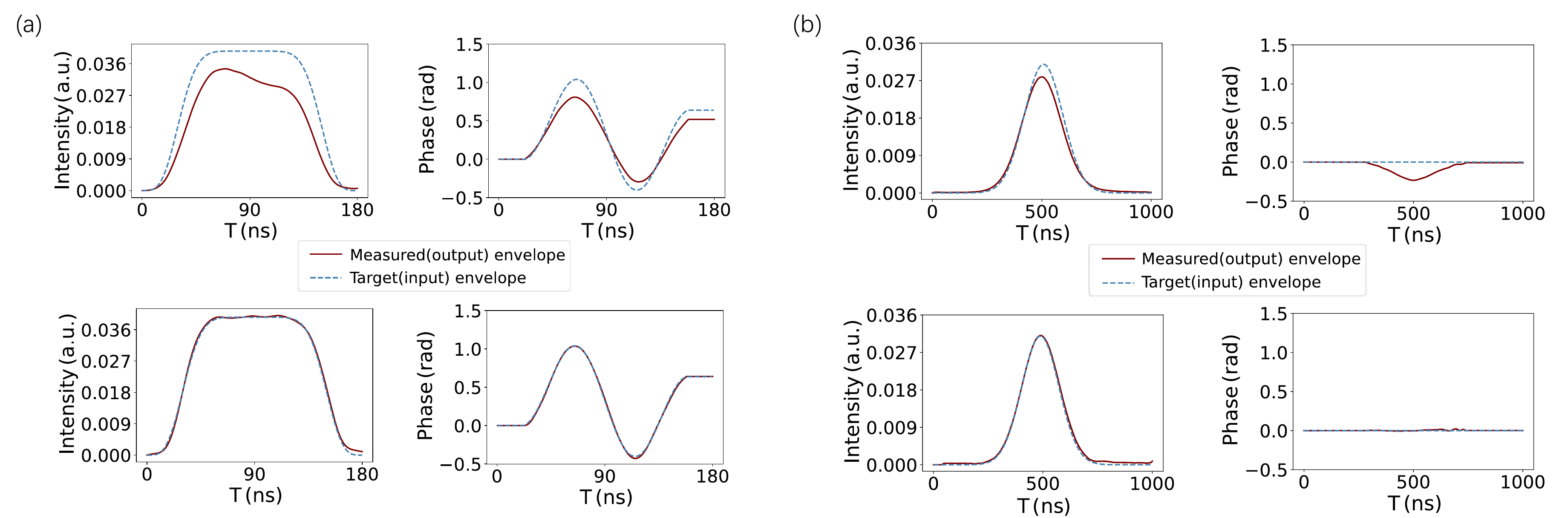}
\caption{\label{fig:experimental_result}{The uncorrected and corrected amplitude and phase of the laser pulse shape. In the upper row of (a), we show the uncorrected optimal two-qubit gate pulse from Ref.~\cite{jandura2022time} with a 180\,ns pulse duration. The slight dip observed at the peak of the amplitude envelope coincides with the moment where the phase gradient is largest likely exceeding the bandwidth of the optical AWG. The lower row of (a) shows the corrected laser amplitude and phase respectively. In (b), we show both the corrected (upper row) and uncorrected (lower row) 1$\mu s$ Gaussian pulse (pure amplitude modulation) used to implement the STIRAP process. The imperfect correction presumably comes from the shot-to-shot variations in the experiment.}}
\end{figure*}

\noindent
where $n \in \{0, ...,N-1\}$, $N$ is the total number of samples depending on the sampling rate of the oscilloscope. For the estimation of Volterra kernel coefficients, we express the input signal to the system as a matrix $S^{in}$ in which the elements $a_{nk}$ have the form~\cite{singh2023compensating}:

\begin{equation}\label{eq:construct_matrix}
a_{nk} =
\begin{cases}
    1 & \text{if } k = 0 \\
    a_{n-k+1} & \text{if } k \in \{1, 2, \ldots, M\},
\end{cases}
\end{equation}
where $M$, denotes the assumed memory length of the distortion.

Similarly, we can write the output signal as a vector $S^{out}$, and then the relationship of input and output signal in Eq.(\ref{eq:vl}) can be expressed in matrix form:

\begin{equation}\label{eq:construct_matrix}
\vec{S}^{out} = S^{in} \vec{H}_{T},
\end{equation}

\noindent
where $H_{T}$ is the column vector $[h^{0},h^{(1)}_{1},h^{(1)}_{2}...,h^{(1)}_{L-1}]$ to be estimated. 

To estimate the complex-value impulse response function, we solve the matrix equation Eq.~\eqref{eq:construct_matrix} by the method of least squares using the package Krylov.jl in Julia, which is generally used to solve the regularized linear least-squares problem. To avoid the overfitting problem during the process of estimating impulse response function, we carefully select the regularization parameter using the cross-validation method~\cite{lee1965measurement}.

\begin{figure}[htbp]
\centering
\begin{minipage}{0.48\textwidth}
    \footnotesize
    \begin{algorithm}[H]
    \caption{Offline iterations algorithm for pulse correction}
    \label{alg:offline_transfer}
    \SetKwComment{Comment}{// }{}
    \DontPrintSemicolon
    \KwIn{$s_{in}$: input pulse, $s_{out}$: output pulse, $\lambda$: damping}
    \KwOut{Pre-distorted input laser pulse}
    \For{$i = 1 \ldots \text{iterations}$}{
        Fitted\_output = Volterra.filter(est\_impulse\_resp, $S_{in}$)\;
        J = Volterra.jacobian(fitted\_1st, input\_pulse)\;\Comment{compute Jacobian matrix}
        $H = J^{H} * J$\;
        
        $grad = - J * (s^{target} - s^{out})$\;
        
        $s^{pred} = s^{in} - \left( H + \lambda I \right)^{-1} grad$\;\Comment{change $\lambda$ for each iteration}
    }
    \Return{$s^{pred}$}
    \end{algorithm}
\end{minipage}
\hfill
\begin{minipage}{0.48\textwidth}
    \centering
    \includegraphics[height=6.0cm]{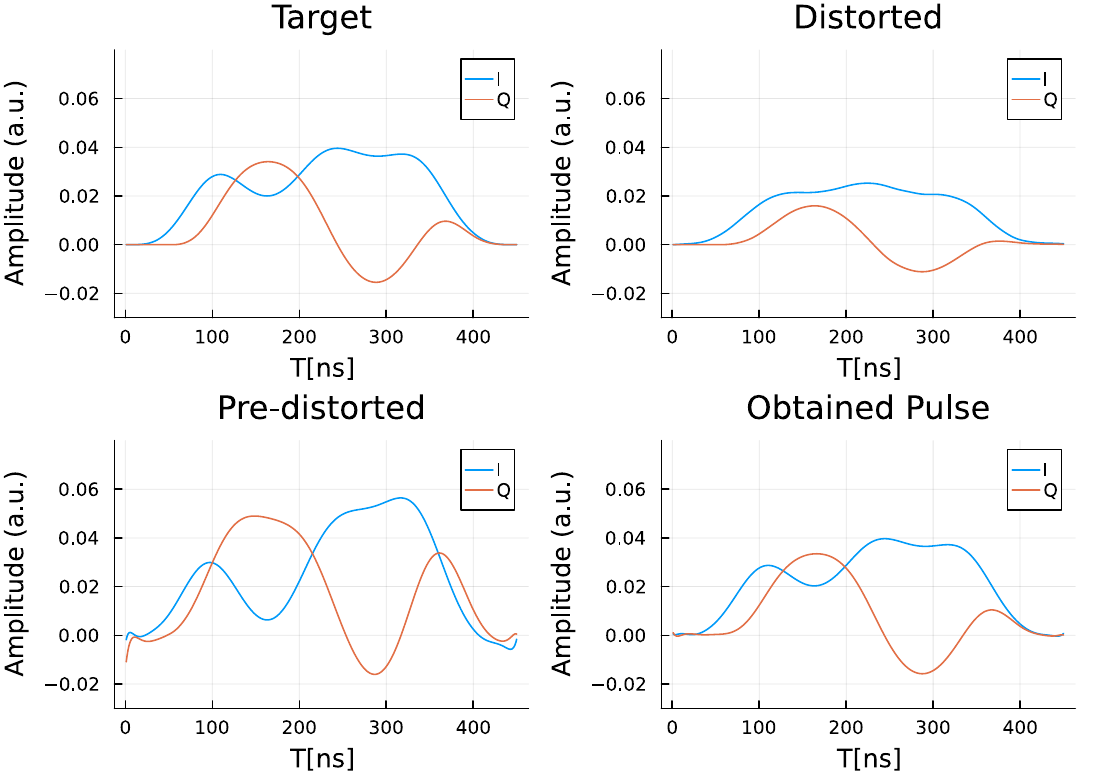}
    \caption{Offline iterations in the transfer function-based method. This method enables us to obtain the pre-distorted imaginary and real parts of $s^{pred}(t)$. We show the target, distorted, pre-distorted, and obtained (output) pulse of the I(t) and Q(t) of the optimal CZ gate. Once the cost function between the target (input) and obtained pulse falls below $1 \times 10^{-6}$, we utilize the corresponding pre-distorted pulse as the new input pulse, integrating it into the closed-loop feedback.}
    \label{fig:offline}
\end{minipage}
\label{fig:combined_figure}
\end{figure}

Once we have successfully estimated the impulse response function of the system, we can include this information in the Levenberg-Marquardt algorithm-based optimization~\cite{more2006levenberg} to generate the pre-distorted $I^{pred}(t)$ and $Q^{pred}(t)$, 
the pre-distorted pulse $s^{pred}(t) = I^{pred}(t) + iQ^{pred}(t)$ is then given by

\begin{equation}\label{eq:predist}
s^{pred}(t) = s^{in}(t) - \left( H + \lambda I \right)^{-1} \nabla \mathcal{C}.
\end{equation}
A proper estimation of $s^{pred}(t)$ can be obtained via iteration whereby $s^{in}(t)$ can be substituted by $s^{pred}(t)$ from the previous iteration.
The cost function, denoted as 
$\mathcal{C}$, is defined as the mean squared error (MSE) between the target and output signals. $\nabla \mathcal{C}$ is estimated from the difference between target and output waveforms pre-multiplied by the Jacobian matrix constructed from the elements of the estimated $H_{T}$~\cite{al2009computing}. The offline iterations algorithm, as shown in Algorithm~\ref{alg:offline_transfer}, is able to make the fitted output pulse closely match the target pulse. The accuracy of the impulse response function is quantified by the mean absolute scaled error (Eq.~\eqref{eq:MASE}) between the measured output pulses and fitted output pulses. The fitted output pulse is computed using the input pulse and the estimated impulse response function. The value of the damping parameter $\lambda$ needs to be carefully chosen as this parameter both affects the stability of the algorithm and the iteration times it needs to converge on a solution.

We found that the most effective compensation for distortion is achieved by incorporating offline iterations (Fig.~\ref{fig:offline}) into the closed-loop iterations for updating the system's impulse response function, as shown in Fig.~\ref{fig:setup}(b).

In Fig.~\ref{fig:experimental_result}, we present the results of the iterative feedback loop for various pulse shapes with durations of 180\,ns and 1000\,ns. To quantify the relative difference between the measured pulse and the target one, we use the mean absolute scaled error (MASE)~\cite{singh2023compensating}:

\begin{equation}\label{eq:MASE}
\text{MASE} = \frac{1}{N} \sum_{i=1}^N \frac{|s_i^\text{real}|}{||s^\text{real}||} - \frac{|s_i^\text{target}|}{||s^\text{target}||},
\end{equation}

\noindent
where $\|s\|$ is the Euclidean norm of the complex vector.

\begin{figure}[h!]
\centering
\includegraphics[height = 3.7cm]{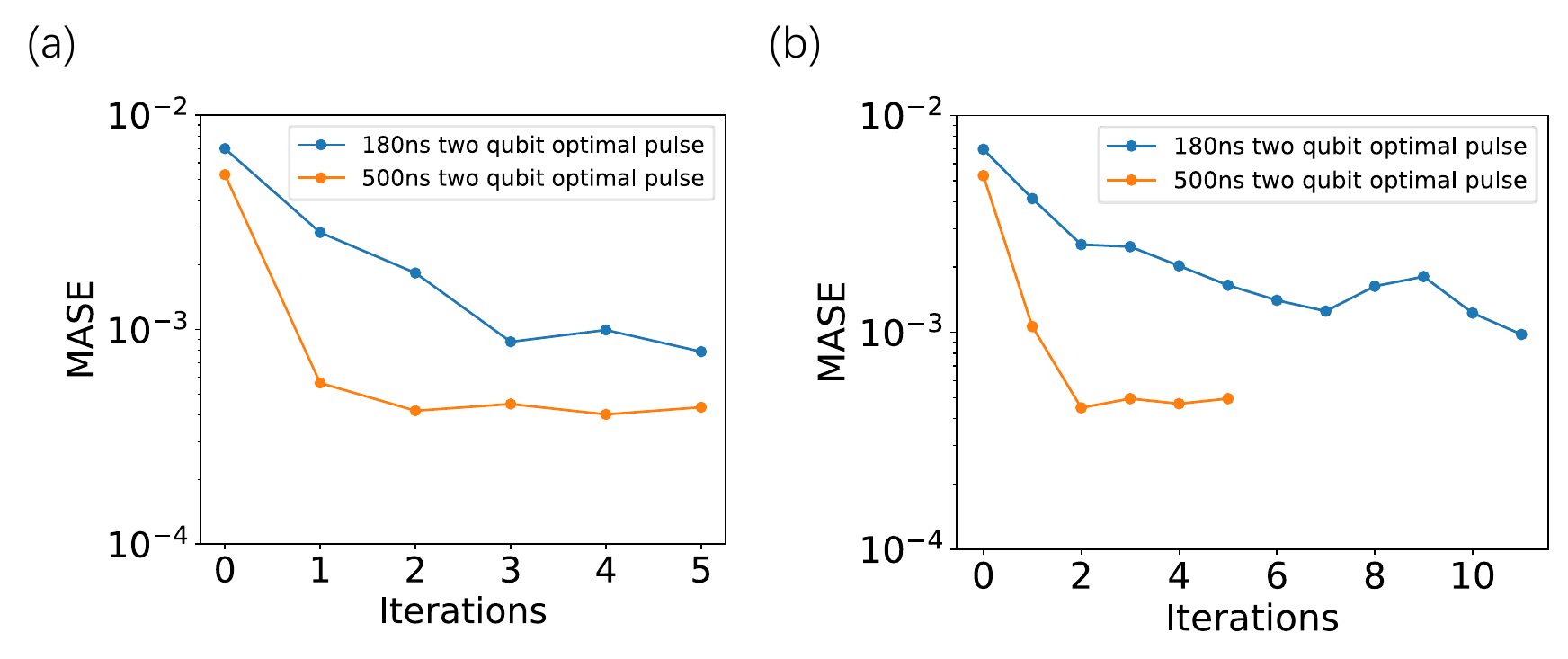}
  \caption{Comparison of iterative closed-loop correction performance: (a) transfer function-based and (b) transfer function-free method for 180\,ns and 500\,ns pulse durations with identical laser pulse envelope shapes (see Fig.~\ref{fig:experimental_result}(a)). The transfer function-free method achieves convergence in under four iterations for the 500\,ns pulse.}
  \label{fig:convergence_time}
\end{figure}

Meanwhile, this closed-loop correction is also compatible with the transfer function-free method~\cite{hansen2003reducing,han2006effect}.  Here, we use the linear error $(s^{target} - s^{out})$ to iteratively update the pre-distorted signal, expressed as $s^{pred} = s^{in} - (s^{target} - s^{out})$. This method, representing a gradient-based optimization with a unity learning rate is computationally efficient, but not numerically stable. In Fig.~\ref{fig:convergence_time}, we compare the performance of the transfer function-based closed-loop correction method with the transfer function-free method for correcting an optimal two-qubit gate laser pulse for different time durations. We notice that both of them only require a few iterations to reach convergence, whereas the transfer function-free method oscillates more around the optimum. However, pulse-to-pulse variations in intensity and phase profiles limit the achievable accuracy, resulting in a residual error of approximately $1 \times 10^{-3}$ for both correction methods.

\section{Conclusion}

In this work, we present an optical AWG capable of generating precisely shaped optical pulses for quantum technology applications. When shaping sub-microsecond pulses we observe significant effects of quadrature distortion attributed to the response of the acousto-optical modulator for a tightly focused beam. To correct for this and other intrinsic pulse distortions, we implement an efficient feedback procedure that involves both online and offline iteration loops, achieving a mean absolute squared error below $10^{-3}$ for pulses down to 180 ns in duration. Our optical AWG is particularly well suited for fast phase and amplitude-controlled pulses desired for time-optical and robust quantum control. Moreover, the compact and robust design can also be readily adapted for rack-based optical systems and for scaling to multiple independent channels for parallel qubit control.

\section*{Acknowledgements}
This research has received funding from the European Union’s Horizon 2020 research and innovation programme under the Marie Skłodowska-Curie grant agreement number 955479. This work has benefited from a state grant managed by the French National Research Agency under the Investments of the Future Program with the reference ANR-21-ESRE-0032 ``aQCess - Atomic Quantum Computing as a Service'', the Horizon Europe programme HORIZON-CL4-2021-DIGITAL-EMERGING-01-30 via the project ``EuRyQa - European infrastructure for Rydberg Quantum Computing'' grant agreement number 10107014 and support from the Institut Universitare de France (IUF). It is also supported by the Interdisciplinay Thematic Institute QMat under the framework of the French Investments for the Future Program (ANR 10 IDEX 0002, ANR 20 SFRI 0012, EUR QMAT ANR-17-EURE-0024).


\nopagebreak[4]
\appendix

\begin{appendices}

\begin{center}

\begin{figure*}
\centering
\includegraphics[width=0.9\linewidth]{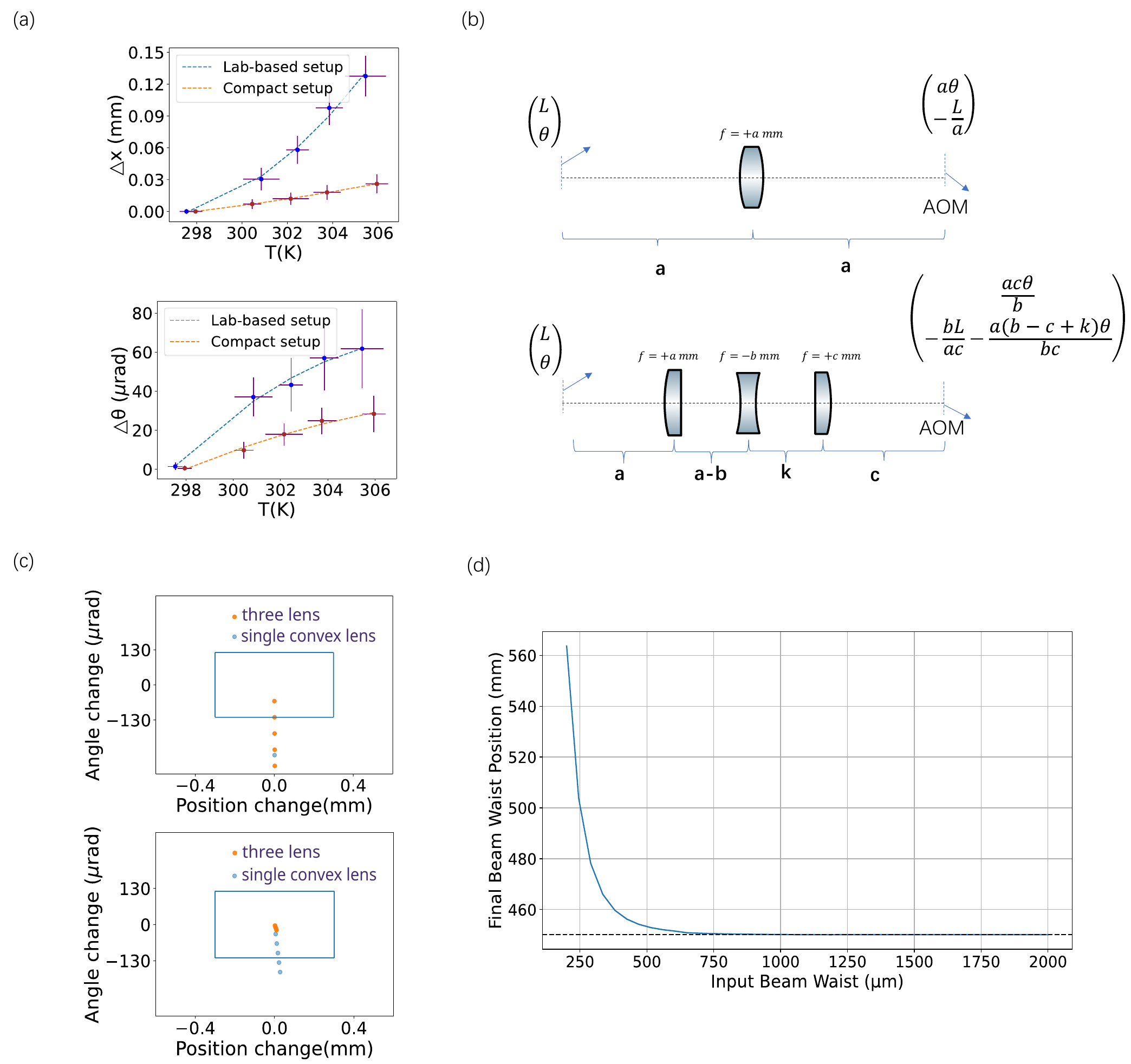}
\caption{(a) Robustness comparison of lab-based and compact setups under varying environmental temperatures. Data points represent averages of 5 measurements at each temperature, with error bars showing standard error. (b) The ABCD matrix analysis of optical configurations of the convex lens(top) and the 3-lens(bottom). Three-lens configuration, which consists of one Galileo telescope and a convex lens. (c) The performance of single lens and three-lens optics configuration integrated in lab-based setup and compact setup respectively, the blue square indicates the active aperture of AOM being 0.6mm and the tolerance of Bragg angle being $125\,\mathrm{\mu rad}$. (d) The beam waist position, coinciding with the AOM location, varies in relation to the input laser beam's waist characteristics in our compact setup. The black dot line indicates the coordinate position of AOM in our setup.}
\label{appendix_fig}
\end{figure*}

\section*{APPENDIX A}

\end{center}

Here, we present a compact design optical modulator with optics mounts (Thorlabs Polaris series) ensuring long-term stability.

We first quantify the pointing accuracy at the position of AOM after the laser beam travels through optical elements in both laboratory-based and compact setups. To evaluate temperature-induced angular and positional deviations, we employ a Beam Profiler to measure the beam's coordinates at two distinct points along the propagation axis under varying temperature conditions, allowing us to infer the change in position and angle of the laser beam at the position of AOM, as shown in Fig.~\ref{appendix_fig}(a).

To optimize laser pulse performance with minimized rise times, we implement a three-lens configuration consisting of a Galilean telescope and a convex lens, rather than a single convex lens. We evaluate the robustness of this configuration against a single convex lens using ABCD matrix calculations within the ray optics framework (Fig.~\ref{appendix_fig}(b)). The initial ray coordinate is denoted as $(0,0)^T$, while $(L,\theta)^T$ represents the new coordinate due to temperature induced deviations. Our analysis demonstrates that the three-lens configuration further mitigates angle deviations of the laser beam at the AOM position, improving beam stability and precision. One can also carefully adjust the distance $k$ shown in the Fig.~\ref{appendix_fig} to make three lens more robust against angle deviations. Furthermore, our simulations (Fig.~\ref{appendix_fig}(c)) of the ray's final coordinate relative to the AOM's active aperture show that, in the compact setup, the three-lens configuration exhibits enhanced robustness against deviations and ensures beam concentration within the AOM's working region. In Fig.~\ref{appendix_fig}(d), we also show that maintaining an input laser beam waist exceeding 750 \,$\mu m$ ensures optimal overlap between the beam waist and the AOM crystal in our compact setup.

\end{appendices}

\newpage

\bibliography{Reference}

\end{document}